\documentclass[%
 aip,
 amsmath,amssymb,
 reprint,%
]{revtex4-1}

\usepackage{graphicx}
\usepackage{dcolumn}
\usepackage{bm}

\usepackage[utf8]{inputenc}
\usepackage[T1]{fontenc}
\usepackage{mathptmx}
\usepackage{etoolbox}

\usepackage{xcolor}
\usepackage{filecontents}
\usepackage{gensymb}
\usepackage{comment}
\usepackage{soul,xcolor}

\makeatletter
\def\@email#1#2{%
 \endgroup
 \patchcmd{\titleblock@produce}
  {\frontmatter@RRAPformat}
  {\frontmatter@RRAPformat{\produce@RRAP{*#1\href{mailto:#2}{#2}}}\frontmatter@RRAPformat}
  {}{}
}%
\makeatother
\begin{document}

\preprint{AIP/123-QED}

\title{$1/f$ noise and two-level systems in MBE-grown Al thin films
}

\author{Shouray Kumar Sahu}
\affiliation{International College of Semiconductor Technology, National Yang Ming Chiao Tung University, Hsinchu 
30010, Taiwan}

\author{Yen-Hsun Glen Lin}%
\affiliation{Graduate Institute of Applied Physics and Department of Physics, National Taiwan University, Taipei 106319, Taiwan}

\author{Kuan-Hui Lai}
\affiliation{Graduate Institute of Applied Physics and Department of Physics, National Taiwan University, Taipei 106319, Taiwan}

\author{Chao-Kai Cheng}
\affiliation{Graduate Institute of Applied Physics and Department of Physics, National Taiwan University, Taipei 106319, Taiwan}

\author{Chun-Wei Wu}
\affiliation{Department of Electrophysics, National Yang Ming Chiao Tung University, Hsinchu 30010, Taiwan}

\author{Elica Anne Heredia}
\affiliation{International College of Semiconductor Technology, National Yang Ming Chiao Tung University, Hsinchu 30010, Taiwan}

\author{Ray-Tai Wang}
\affiliation{Department of Electrophysics, National Yang Ming Chiao Tung University, Hsinchu 30010, Taiwan}

\author{Yen-Hsiang Lin}
\affiliation{Department of Physics, National Tsing Hua University, Hsinchu 300044, Taiwan}

\author{Juainai Kwo}
\affiliation{Department of Physics, National Tsing Hua University, Hsinchu 300044, Taiwan}

\author{Minghwei Hong}
\affiliation{Graduate Institute of Applied Physics and Department of Physics, National Taiwan University, Taipei 106319, Taiwan}

\author{Juhn-Jong Lin}
\altaffiliation [Authors to whom correspondence should be addressed: ] {jjlin@nycu.edu.tw (J.J.L.) and ssyeh@nycu.edu.tw (S.S.Y.)}

\affiliation{Department of Electrophysics, National Yang Ming Chiao Tung University, Hsinchu 30010, Taiwan}

\author{Sheng-Shiuan Yeh}

\altaffiliation [Authors to whom correspondence should be addressed: ] {jjlin@nycu.edu.tw (J.J.L.) and ssyeh@nycu.edu.tw (S.S.Y.)}

\affiliation{International College of Semiconductor Technology, National Yang Ming Chiao Tung University, Hsinchu 30010, Taiwan}
\affiliation{Center for Emergent Functional Matter Science, National Yang Ming Chiao Tung University, Hsinchu 30010, Taiwan}

\date{\today}

\begin{abstract}

Aluminum thin films are essential to the functionalities of electronic and quantum devices, where two-level systems (TLS) can degrade device performance. MBE-grown Al films may appeal to these applications due to their low TLS  densities. We studied the energy distributions of TLS densities, $g(E)$, in 10-nm-thick MBE-grown and electron-beam evaporated Al films through $1/f$ noise measurements between 80 and 360 K. At 300 K, the noise magnitudes in MBE-grown films are about three times lower than in the electron-beam evaporated films, corresponding to the $g(E)$ values about ten times lower in the former than in the latter. Compared with previously established observations, we identified that the $1/f$ noise was generated by thermally activated TLS at grain boundaries.
\end{abstract}

\maketitle

\section{Introduction}
Aluminum thin films play a crucial role in electronic and quantum devices. They have been widely used as a material to enable high-mobility transistors based on the two-dimensional material black phosphorus.\cite{Perello2015} In addition, they have served as a low-contact-resistivity metal for a wide range of semiconductors, including GaAs, GaN, and ZnSe.\cite{Ragay1993, Foresi1993, Janssen2020} Moreover, Al thin films are one of the primary materials for fabricating superconducting qubits.\cite{Place2021} Ultra-thin Al films have been utilized for parametric amplification \cite{Chien2023} and the kinemons, {\em i.e.}, inductively shunted transmon artificial atoms. \cite{Kalacheva2024} In all these applications, two-level systems (TLS) embedded within Al films that are associated with the device operation frequency are known to degrade device performance.\cite{Dutta1979, Paladino2014, McRae2020, Falci2024} The defect densities in Al films grown by molecular beam epitaxy (MBE) can be significantly reduced,\cite{Tournet2016} which can simultaneously minimize the number density of TLS.\cite{Paladino2014} MBE-grown Al superconducting resonators fabricated on the smooth and clean surfaces of single-crystal sapphire substrates exhibit internal quality factors ($Q_i$) approaching $2\times 10^6$ for single-photon excitations in the resonator.\cite{Megrant2012} However, quantitative information on TLS in MBE-grown Al films is still lacking. 

It is established that TLS in metals cause low-frequency $1/f$ resistance noise, where $f$ is the frequency.\cite{Dutta1979, Fleetwood2015} The energy distribution of TLS density per unit energy and per unit volume, $g(E)$, can thus be extracted from the measured $1/f$ resistance noise and sample resistivity ($\rho$). \cite{Pelz1987,Yeh2018} In this work, we have studied the $1/f$ noise in 10-nm-think MBE-grown Al films and extracted $g(E)$ values. Two electron-beam evaporated Al films of the same thickness were also studied for comparison. 

\section{Experimental Method}
The detailed growth method of MBE-grown Al films was described previously.\cite{Lin2024} A 10-nm-thick Al layer was first grown on a sapphire substrate through MBE. Subsequently, a 3-nm-thick Al$_2$O$_3$ capping layer was deposited via electron-beam evaporation. The deposition processes were carried out in an ultra-high-vacuum (UHV) of around $1 \times 10^{-10}$ Torr or lower to prevent any undesirable oxidation and the adsorption of impurities. This growth method prevented any formation of uncontrolled native oxides, such as hydroxides, on the top surface of the Al film. It also effectively reduced potential crystal defects in the oxide layer.\cite{Murray2021, Lin2024} The capped Al film was etched using a BCl$_3$ and Cl$_2$ gas mixture in an inductively coupled plasma-reactive ion etching (ICP-RIE) system to form a multiple-electrode geometry, as shown in Fig. 1(a). For the electron-beam evaporated films grown on 300-nm SiO$_2$ capped Si wafer, we employed the standard electron-beam lithography technique to pattern a similar geometry. 

Table I lists the relevant parameters for all the films studied. The first letter of the sample name, M (P), stands for the MBE-grown (electron-beam-evaporated) film. Figure~1(a) depicts our AC resistance bridge circuit for the $1/f$ resistance noise measurements.\cite{Scofield1987} A modulation and demodulation technique 
was employed in this setup. The circuit contained a preamplifier (Stanford Research Systems Model SR560), a lock-in amplifier (Stanford Research Systems Model SR830), and a dynamic signal analyzer (Stanford Research Systems Model SR785), as described elsewhere.\cite{Chiu2017, Yeh2017} 
X-ray diffraction (XRD) was performed using a synchrotron radiation source at beamline BL 17B of the Taiwan Light Source at the National Synchrotron Radiation Research Center (NSRRC). Figure 1(b) shows the XRD scan across the Al (11$\overline{1}$) diffraction peak along the Al [11$\overline{2}$] direction for an M10 film. The lateral grain size was estimated to be $\approx$\,42 nm from the Scherrer equation. We note that the grain size ($\approx$\,9 nm from XRD studies) of our P10 films was much smaller. To evaluate the crystallinity of our MBE-grown Al thin films, we measured the $\theta$-rocking curves and analyzed their full width at half maximum (FWHM). Our 10-nm-thick films exhibited a narrow FWHM of $0.015^{\circ}$, indicating excellent out-of-plane crystalline quality. This value is comparable to or even smaller than those reported in recent studies, where $\theta$-rocking-curve FWHM values ranged from $0.038^{\circ}$ to $0.066^{\circ}$ for Al thin films grown by sputter beam epitaxy\cite{Law2020} or MBE.\cite{Do2024}

\begin{table}
\caption{\label{tableI}
Relevant parameters of 10-nm-thick Al films. $t$ is thickness, $L$ is length, $W$ is width, $\rho$(300\,K) is room-temperature resistivity, and $\rho_0$ is residual resistivity.
}

\begin{ruledtabular}
\begin{tabular}{cccccc}
Sample & $t$ &  $L$ & $W$    & $\rho$(300\,K)   & $\rho_{\rm 0}$   \\ 
 & (nm) &  ($\mu$m) & ($\mu$m)   & ($\mu \Omega$~cm)  & ($\mu \Omega$~cm) \\ 
\hline 
M10A  & 10 & 20  & 3.6  & 4.8  & 1.7   \\
M10B  & 10 & 200 & 4.4  & 5.6  &  -    \\
M10C  & 10 & 12  & 4.0  & 4.4  &  -    \\
P10A  & 10 & 20  & 1.1  & 11   & 7.8   \\
P10B  & 10 & 20  & 1.4  & 9.9  &  -    \\
\end{tabular}
\end{ruledtabular}
\end{table}

\section{Results}

\begin{figure}
	\centering
	\includegraphics[width=1.05 \linewidth]{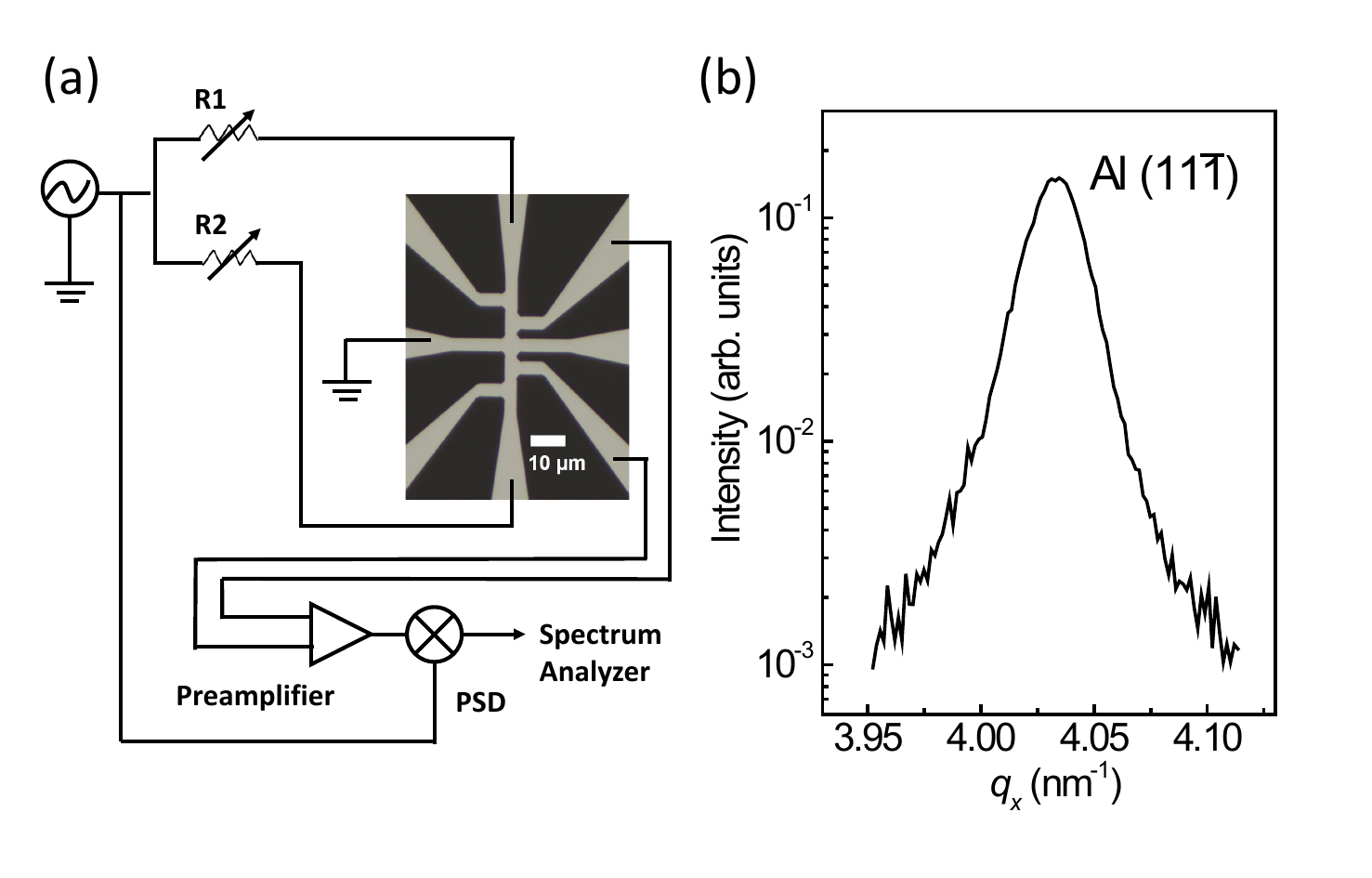}
	\caption{(a) An optical micrograph of a M10 film and a schematic of the $1/f$ noise measurement setup. The ballast resistor $R_1$ was typically a factor $\approx$\,5–10 times greater than the sample resistance. The adjustable resistor $R_2$ was used to balance the bridge. (b) A XRD scan across the Al (11$\overline{1}$) diffraction peak along the Al [11$\overline{2}$] direction for a M10 film. X-axis is the scattering vector $q$, and y-axis is intensity.}
	\label{fig_1}
\end{figure}

\subsection{Resistivity and defect density}

Figure 2(a) shows the temperature ($T$) dependence of resistivity for three M10 and two P10 films. Metallic behavior is observed. The measured resistivity ratios are $\rho$(300\,K)/$\rho_0$ $\approx$\,2.5 and 1.4 in M10A and P10A films, respectively, where $\rho_0 = \rho$(2\,K) is the residual resistivity. We find $\rho_0$(M10A) $\simeq 1.7 ~ \mu \Omega$ cm $\ll$ $\rho_0$(P10A) $\simeq 7.8 ~ \mu \Omega$ cm, indicating significantly fewer defects in the M10 films. 

The average resistivity caused by an individual (static or dynamic) defect is given by $\rho_i \simeq (mv_{\rm F}/ne^2) \times (\sigma_{\rm c}/{\rm vol})$,\cite{Pelz1987, Yeh2017} where $m$, $v_{\rm F}$, $n$, $e$, $\sigma_{\rm c}$, and vol respectively denote the electron effective mass, Fermi velocity, electron density, elementary charge, electron scattering cross-section due to a defect, and the volume of sample. The average scattering cross-section can be approximated by 
$\sigma_{\rm c} \approx 4\pi/k_{\rm F}^2$,\cite{Pelz1987} where $k_{\rm F}$ is the Fermi wavenumber. The residual resistivity is then given through Matthiessen's rule as $\rho_0 = \sum_{i} \rho_i \simeq N_{\rm d} \times \rho_i$, where $N_{\rm d} = n_{\rm d} \times {\rm vol}$ is the total number of defects in the sample, and  $n_{\rm d}$ is the defect density. Given that $m \simeq 0.97\,m_0$ ($m_0$ being the free-electron mass), $v_{\rm F} \simeq 2.1 \times 10^{-6}$ m/s, $k_{\rm F} \simeq 1.7 \times 10^{10}$ m$^{-1}$, and $n \simeq 1.8 \times 10^{29}$ m$^{-3}$ in Al and using the measured $\rho_0$ values, we evaluate $n_{\rm d} \simeq 1.0 \times 10^{27}$ m$^{-1}$ and $4.8 \times 10^{27}$ m$^{-3}$ in M10A and P10A films, respectively. Thus, $n_{\rm d}$ in our MBE-grown films is $\sim$\,5 times lower than that in the electron-beam evaporated films. We have used the bulk material parameters for the above evaluations, because our film thickness is much larger than both the lattice constant ($\approx 0.4$~nm) and the Fermi wavelength $2 \pi/k_F \approx 0.4$~nm of Al.  In the following, we turn to study the fraction of $n_{\rm d}$ which is dynamic ({\it i.e.}, the TLS) and hence contributes to generating 1/$f$ noise in the sample.

\subsection{Low-frequency noise}

\begin{figure}
    \centering
    \includegraphics[width=1.05 \linewidth]{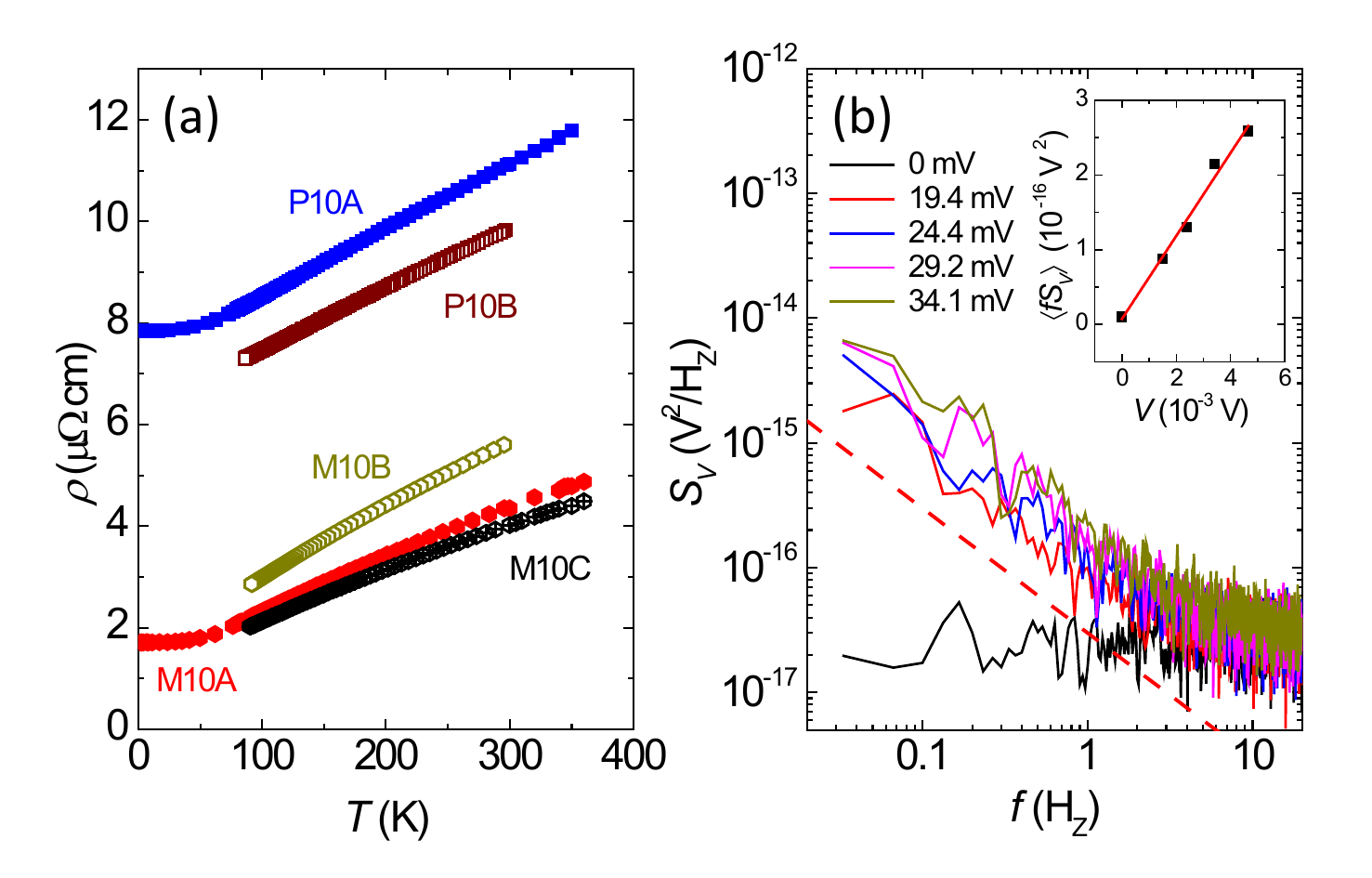}
    \caption{(a) Resistivity as a function of temperature for five Al films. The lowest temperature shown here is 2~K. The M10A and P10A films exhibit a superconducting transition at 1.3 and 1.7~K, respectively (not shown). (b) PSD of M10A film at 300 K under various bias voltages $V_{\rm rms}$. The dashed line indicates $S_V \propto f^{-1}$ and is a guide to the eye. In the AC resistance bridge measurement scheme [Fig. 1(a)], the total bias voltage drop across the sample  $V=2V_{\rm rms}$ [Eq. (1)], where $V_{\rm rms}$ is the root-mean-square voltage drop across one-half of the sample.\cite{Scofield1987} The inset shows $\left \langle f S_V \right \rangle \propto V^2$. The straight solid line is a linear fit.}
    \label{fig_2}
\end{figure}
 
For an ohmic conductor under a small bias current $I$, the $1/f$ resistance noise can be converted into $1/f$ voltage noise, and the measured voltage noise power spectrum density (PSD) is given by the empirical expression \cite{HOOGE1969} 
\begin{equation}
S_V=\frac{\gamma}{N_c f^{\alpha}}V^2+S_V^{0}\,,
\label{eq:Hooge}
\end{equation}
where $\gamma$ is the Hooge parameter which characterizes the noise magnitude, $N_c = n \times \text{vol}$ is the total number of charge carriers in the sample, $V$ is the bias voltage drop across the sample, and $S_V^0$ is the background noise. Figure 2(b) shows the $f$ dependence of $S_V$ at several $V$ values for the M10A film at $T$ = 300 K. The background noise in our circuit was $S_V^{0} \approx 2 \times 10^{-17}$ V$^2$/Hz, which is dominated by the input noise of the SR560 preamplifier. At low frequencies, we find $S_V$ increasing with decreasing $f$, obeying a dependence $S_V \propto f^{-\alpha}$, with an exponent $\alpha \approx 1$ for all bias voltages. To obtain an accurate $\gamma$ value (in the $f^{-1}$ frequency regime), we take the average $\left \langle f S_V \right \rangle = \gamma V^2/N_c + \left \langle f \right \rangle S_V^{0}$ from Eq. (1), where $\left \langle f S_V \right \rangle$ is the mean value of the product of each discrete $f_i$ and $S_{V_i}$ in the data set, and $\left \langle f \right \rangle$ is the mean frequency of $f_i$, where $i$ denotes the $i$th data point. 
The slope $\gamma/N_c$ of a $\left \langle f S_V \right \rangle$ versus $V^2$ plot then gives the $\gamma$ value, see the inset of Fig. 2(b). 

Figure 3(a) shows our extracted $\gamma$ as a function of $T$ for all films. We see $\gamma$ decreases as $T$ decreases for $T < 300$ K. At 300 K, $\gamma \approx 8 \times 10^{-3}$ and $2 \times 10^{-2}$ in M10A and P10A films, respectively. Note that  $\gamma$(M10A) is about three times smaller than $\gamma$(P10A).  For the M10A, M10C, and P10A films, we have measured the 1/$f$ noise up to 360 K and found a $\gamma$ value peaked at $T_p \approx 320$ K, see Fig. 3(b). This characteristic value of $T_p$ reflects the microscopic properties of TLS contained in the samples.\cite{Koch1985} It has previously been established and identified that, in Al films, a value $T_p \approx 325$ K resulted from the diffusion of Al atoms (or, equivalently, vacancies) along grain boundaries.\cite{Koch1985} Our observation of a $T_p \approx 320$~K is in close agreement with this result. A similar $T_p$ value ($\approx 323$ K) in Al films was found by Cottle and Chen.\cite{Cottle1988} An assertion that the 1/$f$ noise in Al films arose from the diffusion of defect Al atoms along grain boundaries was also reported by Smith {\it et al.}\cite{Smith1994} In addition to Al films, grain-boundary generated $1/f$ noise was found in, for example, thin Au films,\cite{Verbruggen1987}, indium tin oxide (ITO) films,\cite{Yeh2013} and graphene.\cite{Kochat2016} [When $T$ is lowered to 100 K, we find $\gamma$(M10A) $\approx 1 \times10^{-3}$ $>$ $\gamma$(P10A) $\approx 3\times 10^{-4}$. This seemingly puzzling inversion will be discussed below.]

\begin{figure}
	\centering
	\includegraphics[width=1.1 \linewidth]{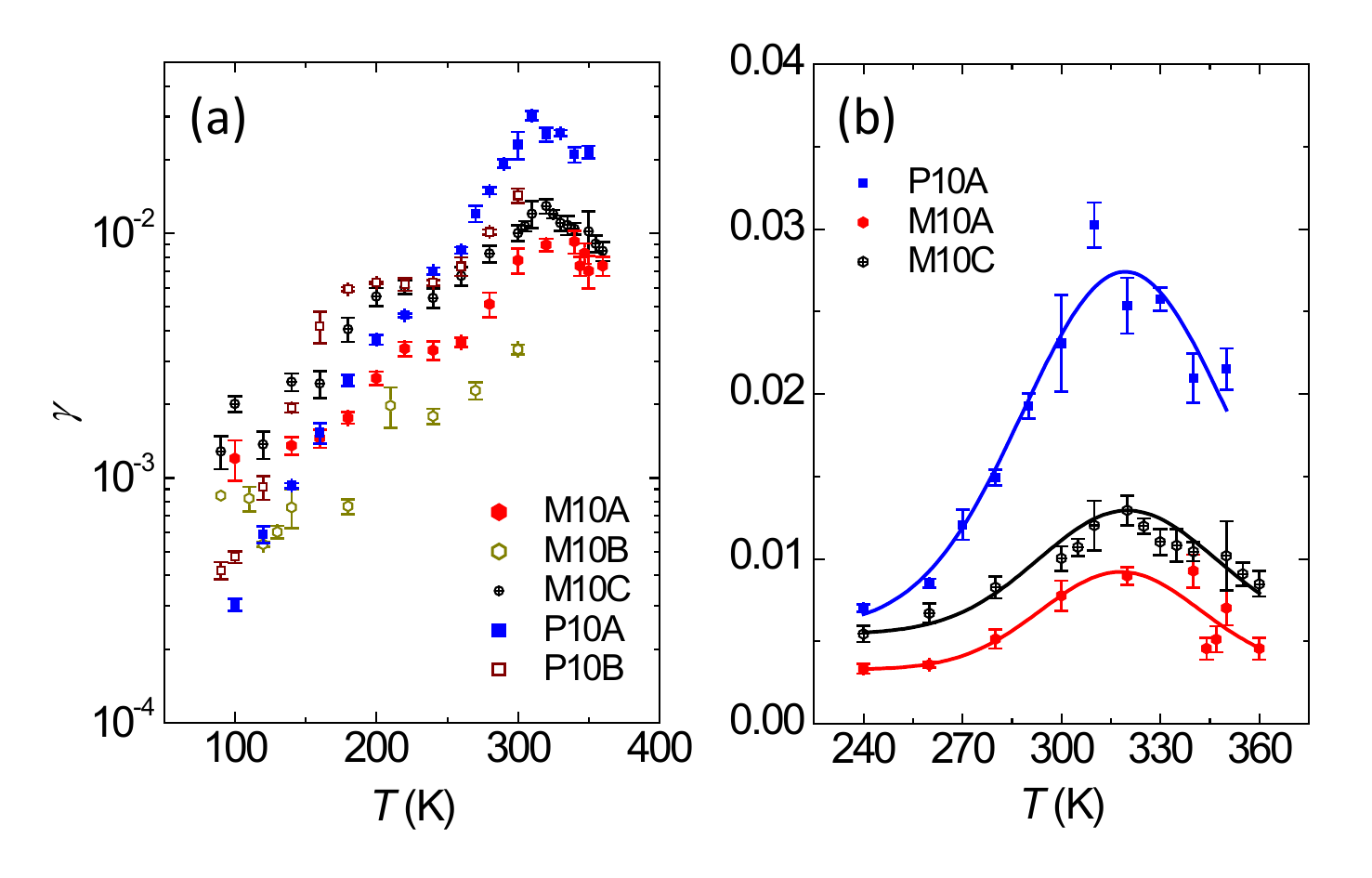}
	\caption{ (a) Variation of $\gamma$ with temperature for five Al films. (b) $\gamma$ versus $T$ for M10A, M10C and P10A between 240 and 360 K. The solid curves are guides to the eye.
}
	\label{fig_1}
\end{figure}

Based on a double-well-potential conception, the Dutta-Dimon-Horn (DDH) model \cite{Dutta1979,Dutta1981} considers an ensemble of independently fluctuating TLS with a broad distribution of relaxation times ($\tau$) or activation energies ($E$). The sum of the resistance noise PSD contributions from all TLS leads to a $1/f^{\alpha}$ dependence. If the distribution of $\tau$ or $E$ is uniform relative to the thermal energy $k_BT$, where $k_B$ is the Boltzmann constant, the theory predicts a value $\alpha = 1$. On the other hand, if the distribution of $\tau$ or $E$ is non-uniform, the theory predicts $\alpha$ deviate (slightly) from 1. Furthermore, if the TLS motion is thermally activated (for example, in contrast to quantum-mechanical tunneling at low temperatures), the theory shows that $\alpha (f,T)$ can be expressed in terms of the normalized noise PSD $S(f,T)$, defined by $S(f,T) = S_V(f,T)/V^2$, through the relation
\begin{equation}
\alpha(f, T)=1-\frac{1}{{\rm ln}(2 \pi f \tau_0)} \left[ \frac{\partial {\rm ln}S(f, T)}{\partial {\rm ln} T} -1 \right], 
\label{eq:exponent}
\end{equation}
where $\tau_0$ has a typical value $\approx$\,$10^{-13}$ s in solids.\cite{Dutta1981}  
Figure 4 shows our measured and calculated $\alpha (f=1\, {\rm Hz}, T)$ values for M10A and P10A films. It is seen that the experimental and theoretical $\alpha$ values agree satisfactorily in both cases. This self-consistency check provides further evidence that the $1/f$ noise originates from thermally activated motions of TLS in Al films at $T >$ 80 K. 

\begin{figure}
	\centering
	\includegraphics[width=0.75 \linewidth]{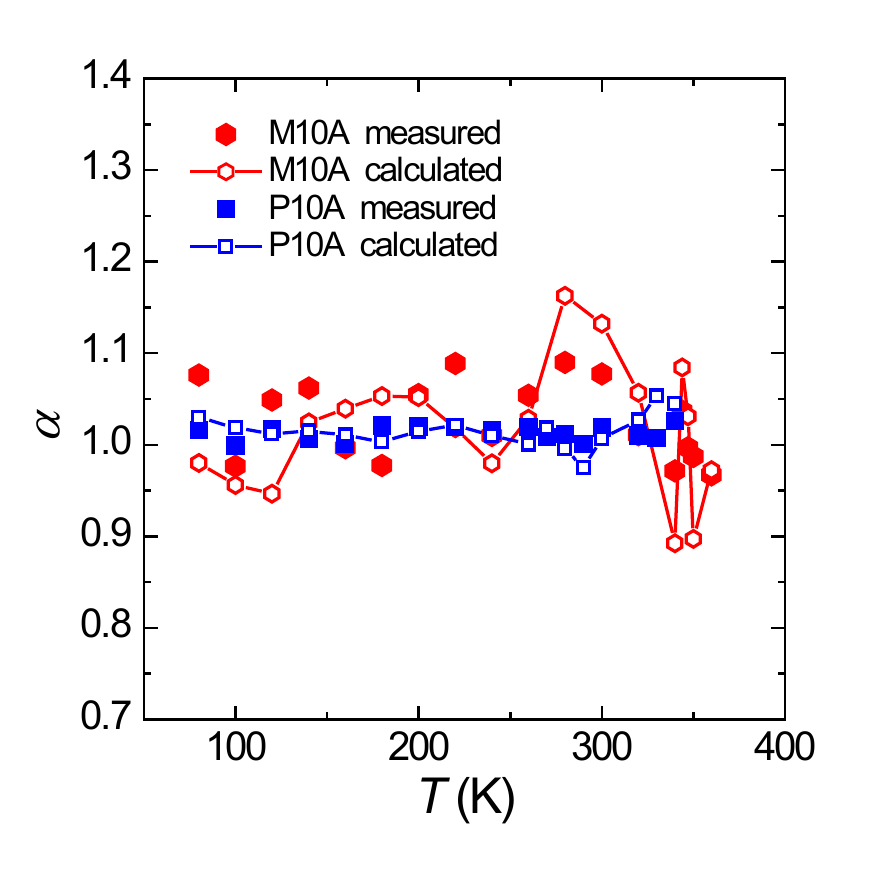}
	\caption{Variation of $\alpha (f = 1\,{\rm Hz})$ with temperature for M10A and P10A films. Theoretical (open symbols) and experimental (closed symbols) values satisfactorily agree. 
    }
	\label{fig_4}
\end{figure}

\subsection{Activation energy distribution} 

\begin{figure}
	\centering
	\includegraphics[width=0.75 \linewidth]{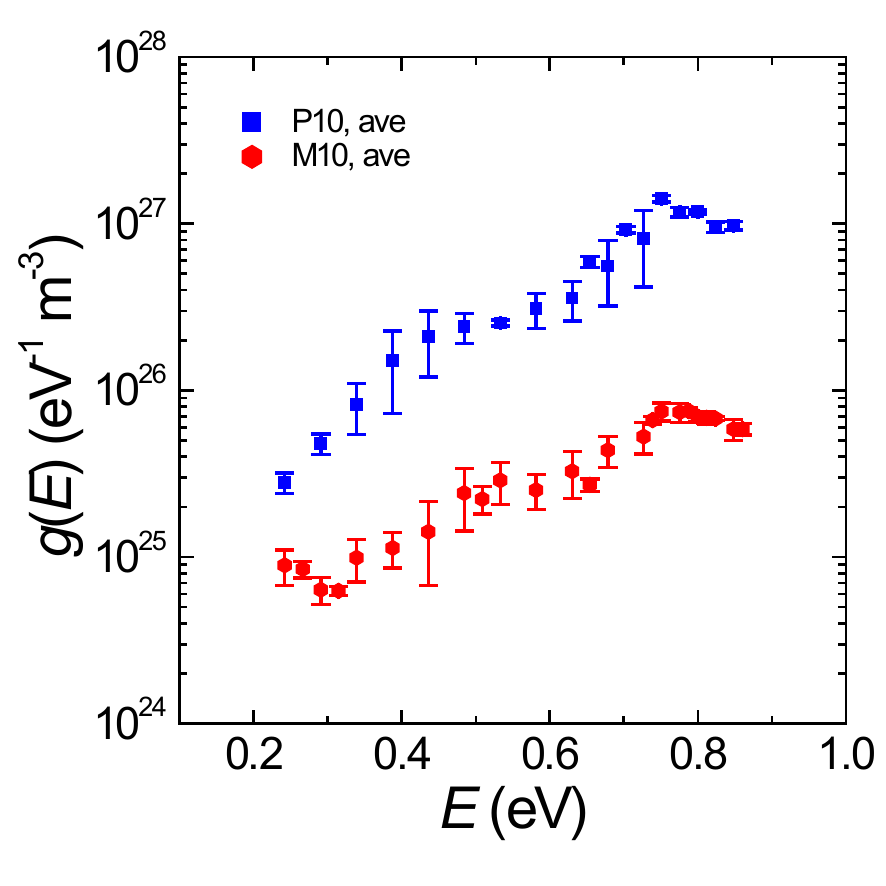}
	\caption{Averaged activation energy distribution $g(E)$ for M10 and P10 films. Red circles are averaged over M10A, M10B, and M10C films, while blue squares are averaged over P10A and P10B films.} 
    \label{fig_5}
\end{figure}

Conceptually, a higher density of TLS will result in a larger resistance noise PSD.\cite{Dutta1979} Under the conditions that the TLS are thermally activated, and combining the scattering theory of Pelz and Clarke,\cite{Pelz1987} we have previously derived an expression for the TLS activation energy distribution $g(E)$ in terms of $\gamma (T)$ and $\rho (T)$ as follows: \cite{Yeh2018}
\begin{equation}
    g(E) \approx \frac{4 \pi n \gamma}{2.6 k_B T}\left [ \frac{\rho e^2}{m v_{\rm F} \sigma_{\rm c}} \right ]^2,
    \label{dsig_1}
\end{equation}
where the TLS activation energy $E$ is given by $E = -k_B T \ln (2 \pi f \tau_0)$. Note that at low frequencies of concern, the product $2\pi f \tau_0 \ll 1$. Figure 5 shows our extracted, averaged $g(E)$ values for the two sets of M10 and P10 films listed in Table 1, where we used $\tau_0 = 10^{-13}$ s and $f$ = 1 Hz to calculate $g(E)$. It is seen that the distributions of $g(E)$ are similar for the two sets of films. They both reveal a maximum at the characteristic energy $E_p \approx 0.78$ eV, with $g(E_p) \approx 7 \times 10^{25}$ ($\approx 1 \times 10^{27}$) eV$^{-1}$ m$^{-3}$ in M10 (P10). This $E_p$ value corresponds to the characteristic temperature $T_p \approx 320$ K discussed above. Figure 5 illustrates that as $E$ decreases from 0.78 to 0.24 eV, $g(E)$ decreases by factors of $\sim$\,8 and $\sim$\,40 in the M10 and P10 films, respectively. Moreover, the $g(E \approx E_p)$ value in the former is about one order of magnitude smaller than in the latter, demonstrating a significantly lower TLS density in M10 films. This supports the attribution of TLS in Al films to the diffusion of Al atoms along grain boundaries. This finding is valuable for present-day prevailing Al-based superconducting and quantum device fabrication. In P10 films, the sharp temperature dependence of $g(E)$ below $T_p$, associated with the weak temperature dependence of $\rho$ due to being relatively disordered, gives rise to a comparatively strong decrease of $\gamma$ [$\propto g(E)/\rho^2$], see Eq.~(3), with decreasing $T$ in Fig.~3(a).

We may further quantitatively estimate the number density of TLS, denoted by $n_{\rm TLS}(T)$, which generates the measured 1/$f$ noise magnitude. The responsible number density is temperature dependent and approximately given by $n_{\rm TLS}(T) \approx g(E) \times 2.6\,k_BT$.\cite{Yeh2018} We obtain at the charactersitic temperature $T_p$ = 320 K the values $n_{\rm TLS}(T_p) \approx 5 \times 10^{24}$ m$^{-3}$ in M10 films and $\approx 8 \times 10^{25}$ m$^{-3}$ in P10 films. These values correspond to dynamic-to-total defect ratios $n_{\rm TLS}/n_{\rm d} \approx 0.5\%$ in the former and $\approx 2\%$ in the latter. These notable ratios indicate that a relatively large fraction of TLS are activated at room temperature.

\textit{Comparison with previous works.} We use Eq.~(3) to calculate  the $g(E)$ values at $T$\,=\,300 K for those Al films studied in Refs.~\onlinecite{Koch1985,Homberg1997}. This temperature corresponds to an activation energy of \(E \approx 0.73\,\text{eV}\). For those 100 nm-thick evaporated polycrystalline films with an average grain size of about 50 nm studied by Koch \textit{et al.}\cite{Koch1985}, we obtain $g[E(T=300\,{\rm K})] \approx 2 \times 10^{25}$ eV$^{-1}$ m$^{-3}$. Homberg \textit{et al.}\cite{Homberg1997} had fabricated 300–450 nm-thick Al films by sputtering followed by guided recrystallization of Al in submicron SiO$_2$ groove patterns. For their bamboo-structured films with a grain size $\sim$\,230 nm, we obtain $g[E(T=300\,{\rm K})] \approx 2 \times 10^{24}$ eV$^{-1}$ m$^{-3}$. For their single-crystal films with a large grain size $>$\,3 $\mu$m, we obtain $g[E(T=300\,{\rm K})] \approx 8 \times10^{23}\,\text{eV}^{-1}\,\text{m}^{-3}$. 
These $g(E)$ values, taken together with our results of Fig. 5, demonstrate a monotonic increase of $g(E)$ with decreasing grain size. Indeed, regardless of the deposition methods, the $g(E)$ values are larger in thinner Al films where grain boundaries are more abundant than in thicker films.

\section{Discussion and Conclusion}

Whether grain-boundary-induced TLS affect the internal quality factor $Q_i$ (Refs. \onlinecite{Megrant2012,Mukhanova2024}) and frequency noise\cite{deGraaf2018,Zeng2025} in the superconducting-resonator regime is an emergent issue. In this work, we have focused only on the 80--360 K temperature range where TLS motions are dominated by thermal activations. At liquid-helium temperatures, quantum tunneling will be responsible for the TLS motions. Moreover, the quantum-interference effect may cause universal conductance fluctuations that may, in turn, enhance the 1/$f$ noise.\cite{Chien2023a} This low-$T$ regime deserves future study. In the literature, it has been widely accepted that $Q_i$ is mainly limited by the TLS at interfaces and/or surfaces.\cite{Earnest2018} On the other hand, a recent experiment on granular-Al superconducting resonators used a series of 91-nm thick films, where the authors varied the oxygen partial pressure during deposition to tune the resistivity $\rho$ values of samples.~\cite{Gupta2025}  It was found that increasing $\rho$, implying increased film granularity due to a thicker inter-grain AlO$_x$ layer, and thus a higher density of grain boundaries, systematically reduced $Q_i$. The authors ascribed the decrease in $Q_i$ to additional losses in their samples. We suspect that grain-boundary TLS are the origin of those additional losses. We should also note that in contrast to relatively thick films (typically 100--200~nm thick) used for superconducting resonators, ultra-thin films are essential for high-kinetic-inductance applications, such as superinductors,\cite{Astafiev2012, Niepce2019} inductance-based qubits,\cite{Kerman2010, Peltonen2018, Grunhaupt2019, Kalacheva2024}, and single-photon detectors,\cite{Day2003} despite their typically higher densities of grain-boundary TLS.     

We have measured the $1/f$ noise of 10-nm-thick MBE-grown and electron-beam evaporated Al films. We found that the $1/f$ noise is governed by thermally activated TLS above 80 K. The noise magnitude $\gamma$ peaks at $\approx$\,320 K, suggesting that the TLS are related to the motion of Al atoms along grain boundaries. At room temperature, the $\gamma$ value in MBE-grown films is about three times lower than in electron-beam-evaporated films, corresponding to an activation energy distribution $g(E)$ value about 10 times lower in the former. Systematic studies of grain-size-dependent effects on resistance noise, the internal quality factor $Q_i$, and frequency noise in superconducting resonators can yield close insight into the roles of grain-boundary TLS.

\begin{acknowledgments}

We thank Pertti Hakonen for fruitful discussion, and Hsien-Wen Wan, Yi-Ting Cheng, and Min-Yen Wu for experimental assistance. This work was supported by the National Science and Technology Council of Taiwan through grant numbers 113-2119-M-007-008 (Y.H.L., M.H., and J.J.L.), 112-2112-M-007-051 (J.K.), 111-2112-M-A49-034 (J.J.L.), and 110-2112-M-A49-033-MY3 and 113-2112-M-A49-036 (S.S.Y.). Y.H.L. acknowledges the support from the Taiwan Ministry of Education Yushan Young Scholar Fellowship. S.S.Y. acknowledges the support from the Taiwan Ministry of Education through the Higher Education Sprout Project of the NYCU.

\end{acknowledgments}

\section*{author declarations}
\noindent\textbf{Conflict of interest} 

The authors have no conflicts to disclose.\\

\noindent\textbf{Author Contributions}

\noindent\textbf{DATA AVAILABILITY STATEMENT}

The data that support the findings of this study are available from the corresponding author upon reasonable request.\\

\noindent\textbf{REFERENCES}


%

\end{document}